\documentclass[11pt,twoside]{article}

\usepackage{asp2006}
\usepackage{epsf}
\usepackage{psfig}
\usepackage{lscape}
\usepackage{graphicx}

\begin{document}
\title{Low-frequency solar p modes as seen by GOLF and GONG instruments}   
\author{D. Salabert}   
\affil{Instituto de Astrof\'isica de Canarias, C/ V\'ia L\'{a}ctea s/n, 38205, La Laguna, Tenerife, Spain}   
\author{R. A. Garc\'ia}
\affil{Laboratoire AIM, CEA/DSM-CNRS, Universit\'e Paris 7 Diderot, IRFU/SAp, Centre de Saclay, 91191, Gif-sur-Yvette, France}

\begin{abstract} 
In the case of spatially-revolved helioseismic data (such as MDI, GONG, HMI), 
the usual mode-fitting analysis consists of fitting the 2$l$+1 
individual $m$-spectra of a given multiplet ($n, l$) either individually 
or simultaneously. Such fitting methods fail to obtain reliable estimates 
of the mode parameters (frequency, splitting, ...) when the signal-to-noise 
ratio (SNR) is low, which makes those methods not suitable when one wants 
to look at the low-amplitude, long-lived solar p modes in the low-frequency 
range. Instead, Salabert et al.~(2007) developed a new method to extract the mode 
parameters by adjusting the rotation- and structure-induced frequency shift for 
each $m$-spectrum to minimize the mode width in the $m$-averaged spectrum 
(a ``collapsogram"). The $m$-averaged spectrum technique, applied to 
the spatially-resolved GONG and MDI data, appeared to be a powerful tool 
for low SNR modes in the low-frequency range. Another possibility to 
increase the SNR is to combine data from different instruments 
(Garc\'\i a et al. 2004a). We present here an adaptation of both techniques: 
the ``collapsograms" applied to a combination of observations from a Sun-as-a-star 
instrument, GOLF, and a disk-imaged one, GONG.
\end{abstract}

\section{Introduction}
After more than 13 years of unprecedented quality helioseismic observations, we have 
not been able to measure acoustic modes below $\sim$~1~mHz 
(e.g. Broomhall et al. 2007) or gravity modes. Indeed, only the asymptotic 
properties of these modes have been uncovered (Garc\'\i a et al. 2007, 2008a), 
but the measurements of individual modes is still a challenge of modern solar 
physics (e.g. Mathur et al. 2007; Appourchaux these proceedings). It will 
be impossible to properly determine the structure and dynamic of the solar 
core without these modes (e.g. Garc\'\i a et al. 2008b; Mathur et al. 2008) 
as well as the high-penetrative, high-order, low-degree p modes (Garc\'\i a, 
Mathur, \& Ballot 2008). In order to progress towards lower frequencies, we 
need to develop new instrumentation --- e.g. the GOLF-NG project (Salabert et al. 
these proceedings), but also to better combine the existent data as well as 
to improve the analyzing techniques. 

As the frequency decreases, the amplitude of the modes decreases 
while the background level increases. This background noise is a combination 
of convective noise, instrumental noise, and, in the case of ground-based 
stations, of the terrestrial atmosphere. Combining data from different instruments 
has the advantage to reduce the instrumental noise as well as the convective 
noise, since each instrument observes at a different height of the solar atmosphere
and the convection is not fully correlated (Garc\'\i a et al. 2004b).

\begin{figure*}[t]
\centering
\begin{tabular}{cc}
\includegraphics[width=0.33\textwidth,angle=-90]{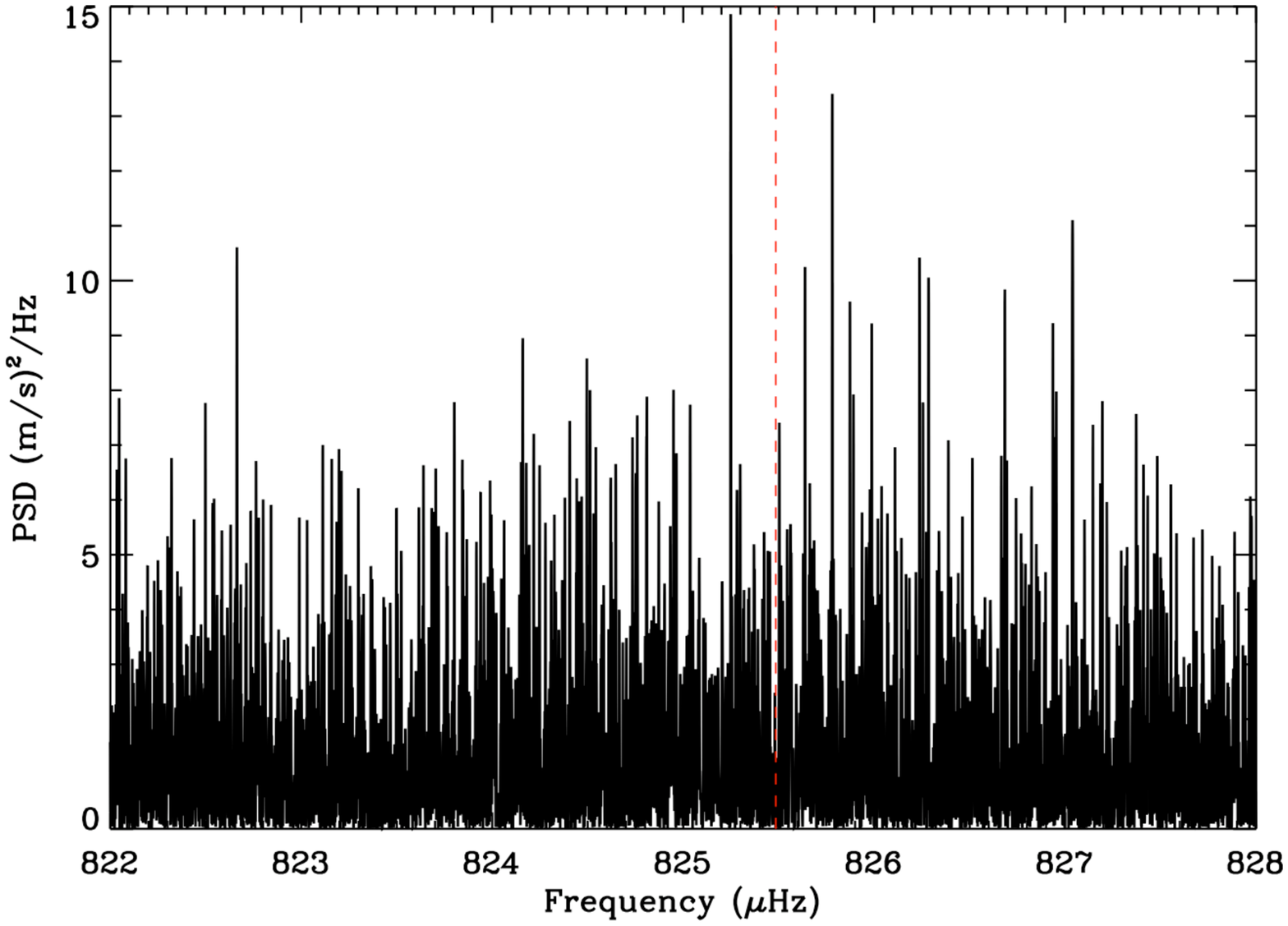} &
\includegraphics[width=0.33\textwidth,angle=-90]{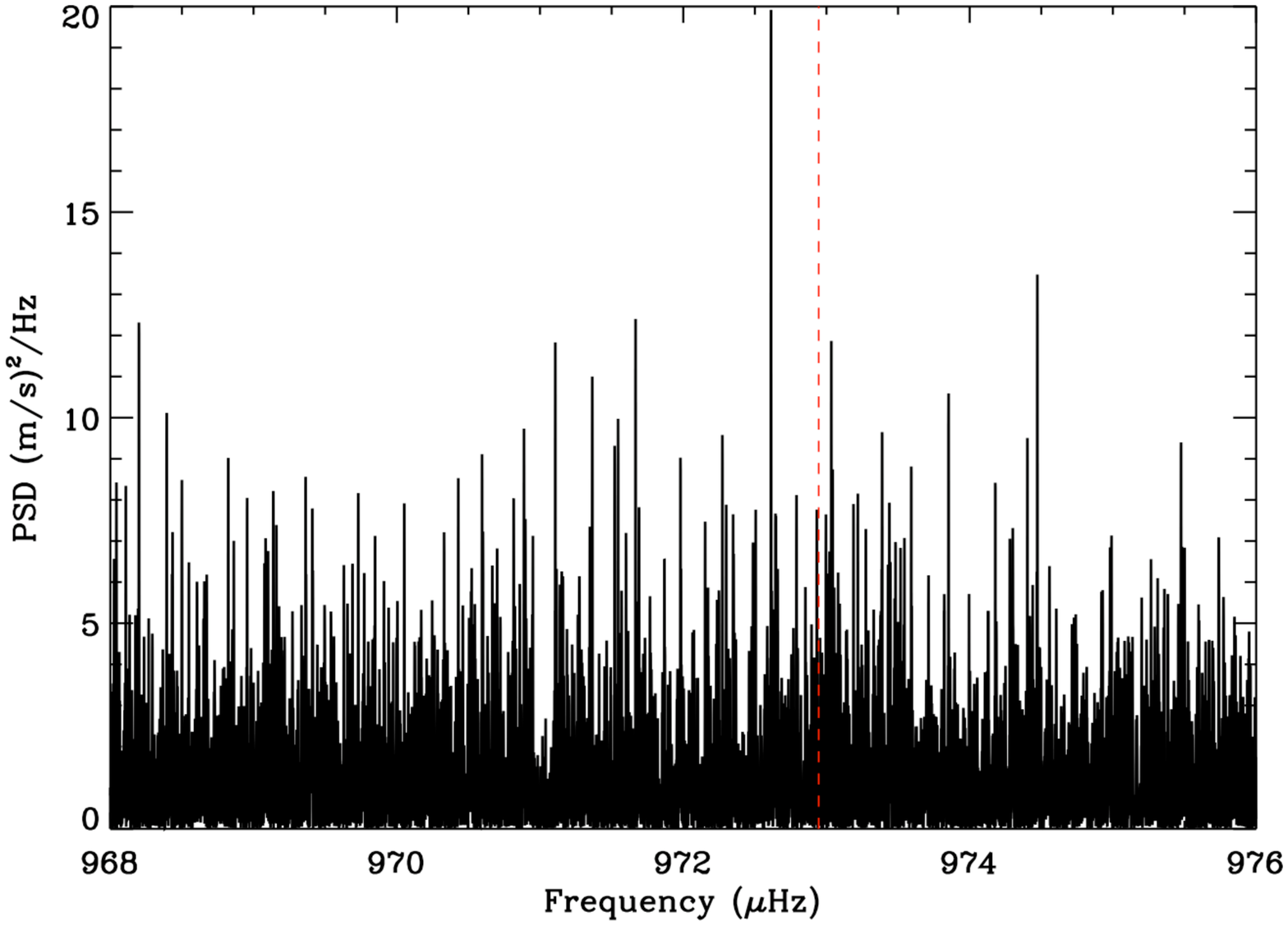} \\
\includegraphics[width=0.33\textwidth,angle=-90]{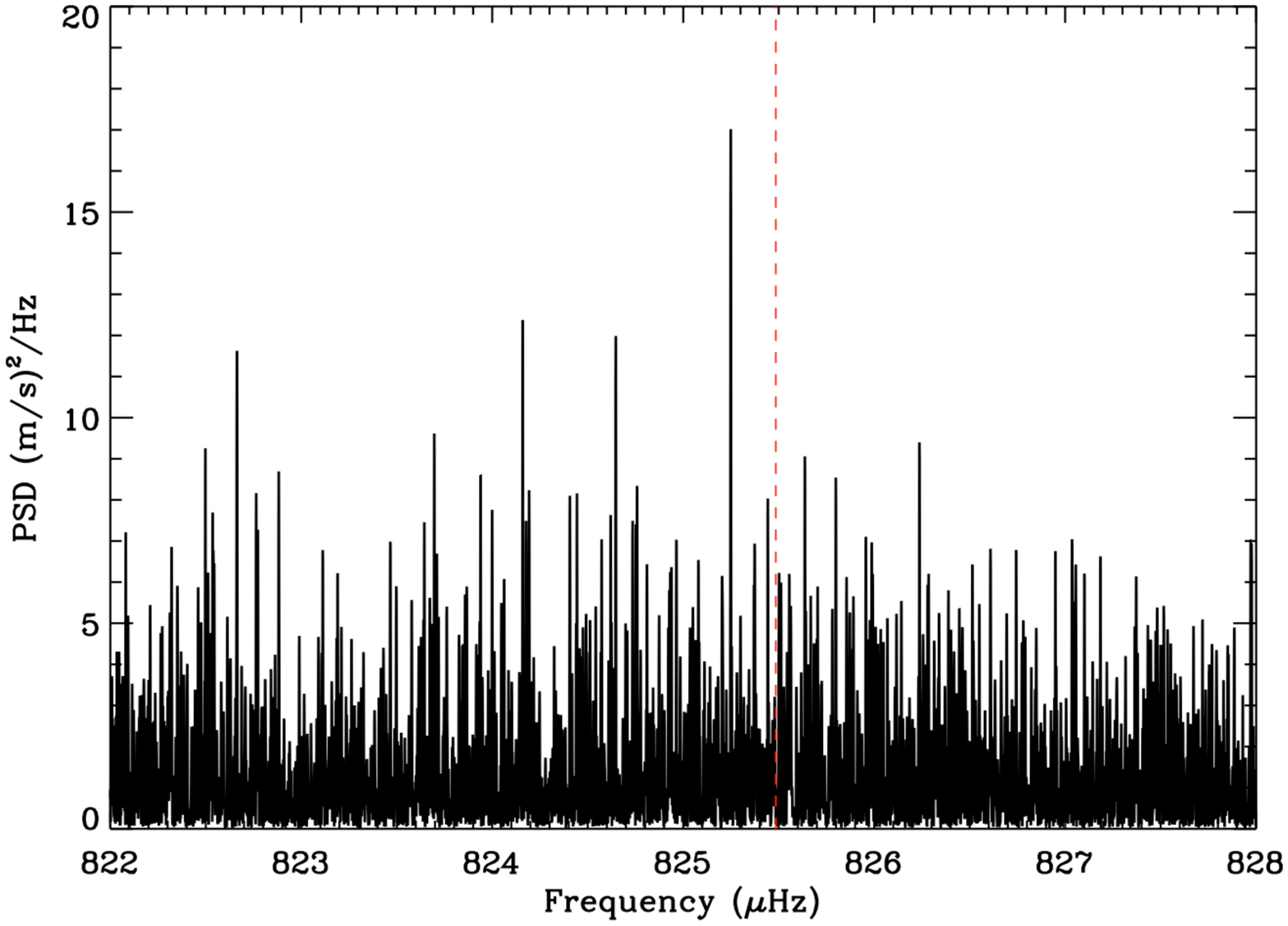} &
\includegraphics[width=0.33\textwidth,angle=-90]{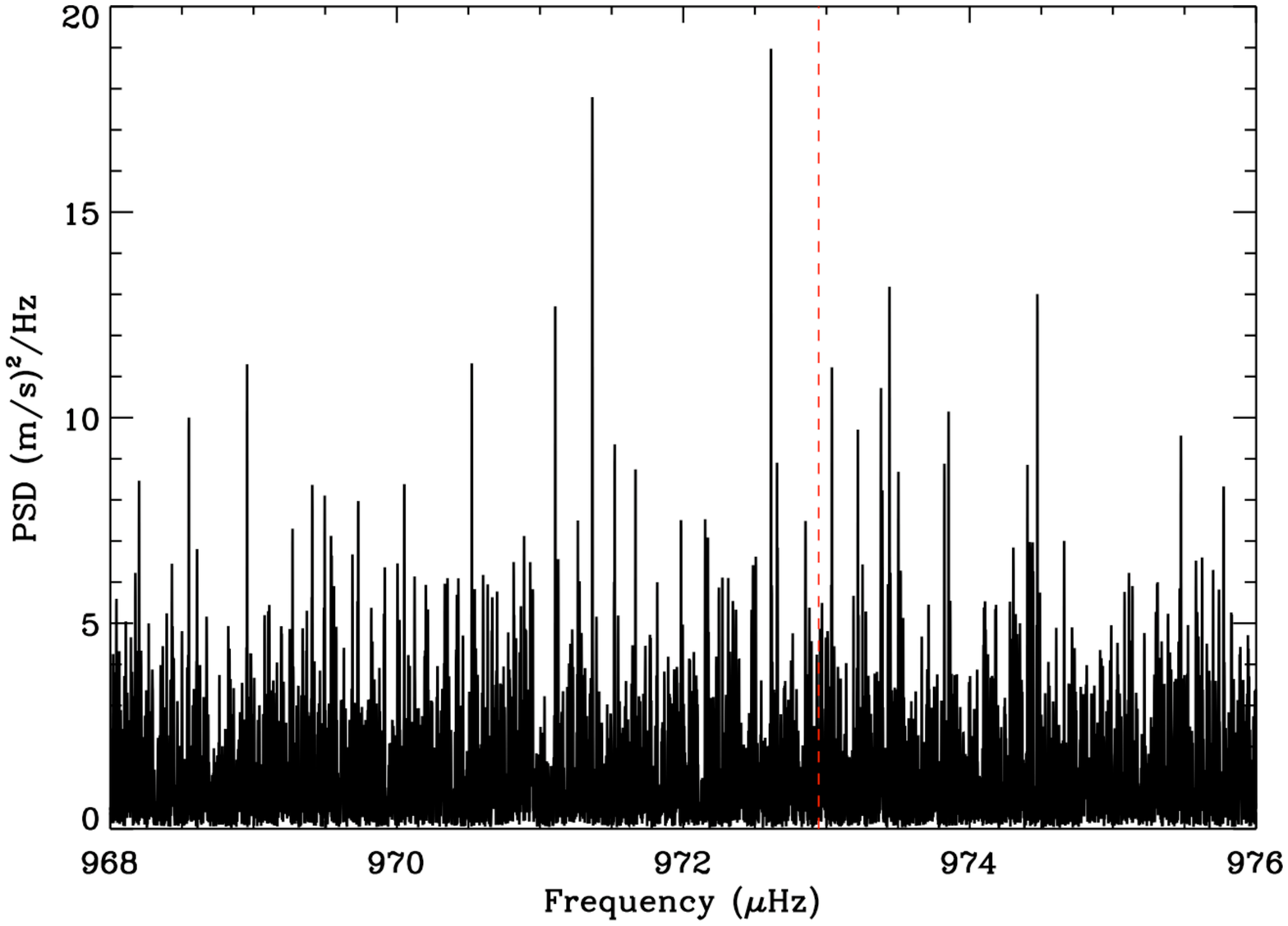} \\
\end{tabular}
\caption{Regions of the power spectrum density (PSD) where the $l=0$, $n=5$ (left panels) and $n=6$ (right panels) are 
expected. The PSD were obtained with the GOLF 4472-day time series (upper panels) and with the 3620-day (GOLF+GONG) 
time series (lower panels). The vertical dashed lines indicate the corresponding theoretical frequency 
from the seismic model (Couvidat, Turck-Chi\`eze \& Kosovichev, 2003)}
\end{figure*}

\section{Data sets and methodology}
In this work, we use 4472-day time series --- from April~11, 
1996 to July~8, 2008 --- calibrated into velocity (Garc\'\i a et al. 2005) 
of the disk-integrated, Sun-as-a-star GOLF/SoHO instrument 
(Gabriel et al. 1995), and 3620-day velocity time series --- from 
April~11, 1996 to March~9, 2006 --- of the spatially-resolved, ground-based GONG 
network (Harvey et al. 1996).
We average the GOLF observations with the GONG time series from each $|m|$-component ($m=0$ and $m=\pm2$) of 
the spherical harmonic projection of the $l=2$ mode, both being filtered by a backwards difference filter (BDF). 
Then, we compute the Fourier transform of each multiplet component with a five times zero padding, 
correct for the effect of the BDF filter (Garc\'\i a \& Ballot 2008) and 
finally average the power spectrum density. The length of the combined 
series is 3620 days.

\section{Power spectra of GOLF and (GOLF+GONG) data sets}
The combined power spectrum does not present the daily harmonics thanks 
to the high duty cycle of the space-based GOLF instrument. Moreover, the 
leakage from high-degree modes is reduced because they are not present in 
the GOLF spectra. The amplitude of the $l=0$ modes increases as they are 
present in both spectra as well as the overall SNR. For example, the mode $l=0$, $n=5$  
has a larger SNR in the combined (GOLF+GONG) data set than in GOLF data alone, even when 
using 20\% shorter time series (see Fig.~1), while the mode $l=0$, $n=6$ has a slightly 
smaller SNR.

\section{Collapsograms applied to the (GOLF+GONG) time series}
\begin{figure}
\begin{center}
\includegraphics[scale=0.64]{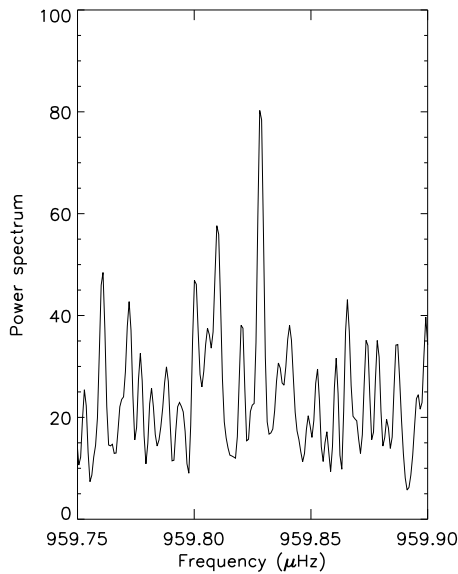} 
\caption{Low-frequency, low-degree $l=2$, $n=5$ mode at $\approx$ 959.83~$\mu$Hz measured with the
collapsogram technique applied to the combined (GOLF+GONG) 3620-day time series. Estimated splitting
$a_1$ = 387.8~nHz.}
\label{fig:colp}
\end{center}
\end{figure}

We applied the collapsogram technique (Salabert et al. 2007) to the $l=2$ combined (GOLF+GONG)
3620-day time series. Several more low-degree modes in the low-frequency range can be then
observed than using the collapsograms applied only to the GONG (or MDI) data, such as the 
$l=2$, $n=5$ mode (Fig.~\ref{fig:colp}). Moreover, the combination of GOLF with GONG
increases the overall SNR, allowing us to measure p modes which would not have a SNR high 
enough to be measured in GOLF only.

\section{Conclusions}
This preliminary analysis shows the advantage of combining observations 
from spatially-resolved helioseismic instruments and disk-integrated, 
Sun-as-a-star instruments. The noise level in the low-frequency range is reduced 
and the SNR of the low-degree modes is increased. Moreover, when combining ground-based data with 
space-based data, there is no signal coming from the daily harmonics. The SNR of the combined 
data is greater than the SNR of only one instrument even with a longer time series. 
With the present (MDI, GONG, GOLF, BiSON) and future (HMI, PICARD, GOLF-NG) 
available helioseismic databases, the combination of observations obtained 
from spatially-resolved and Sun-as-a-star instruments can be of great interest
in the search of the solar low-frequency acoustic and gravity modes. 

\acknowledgements 
This work has been partially founded by the Spanish grant PNAyA2007-62650 
and the CNES/GOLF grant at the SAp-CEA/Saclay. SOHO is an international 
cooperation between ESA and NASA. This work utilizes data obtained by the 
Global Oscillation Network Group (GONG) program, managed by the National Solar
 Observatory, which is operated by AURA, Inc. under a cooperative agreement 
with the National Science Foundation. The data were acquired by instruments 
operated by the Big Bear Solar Observatory, High Altitude Observatory, 
Learmonth Solar Observatory, Udaipur Solar Observatory, Instituto de 
Astrof\'{\i}sica de Canarias, and Cerro Tololo Interamerican Observatory.

\end{document}